\documentclass[aps,prd,nofootinbib]{revtex4}
\usepackage{colordvi}
\usepackage{graphicx,bm}
\usepackage{amsmath,amssymb,mathrsfs,verbatim,subfigure}

\newcommand\Dp{{\Delta_+}}
\newcommand\Dm{{\Delta_-}}
\newcommand\hfrac[2]{{#1}/{#2}}
\newcommand\cO{{\cal O}}
\newcommand\oO{{O}}
\newcommand\deltaT[1]{{\left[#1\right]_T}}
\newcommand\bulk{{\rm bulk}}

\begin{document}

\title{Bulk spectral function sum rule in QCD-like theories with a
  holographic dual}

\author{Paul M. Hohler}
\email{pmhohler@uic.edu}

\affiliation{Department of Physics, University of Illinois,
Chicago, IL 60607-7059, USA}

\author{Mikhail A. Stephanov}
\email{misha@uic.edu}

\affiliation{Department of Physics, University of Illinois,
Chicago, IL 60607-7059, USA}


\begin{abstract}
  We derive the sum rule for the spectral function of the
  stress-energy tensor in the bulk (uniform dilatation) channel in a
  general class of strongly coupled field theories. This class
  includes theories holographically dual to a theory of gravity
  coupled to a single scalar field, representing the operator of the
  scale anomaly. In the
  limit when the operator becomes marginal, the sum rule coincides
  with that in QCD. Using the holographic model, we verify explicitly the
  cancellation between large and small frequency contributions to the
  spectral integral required to satisfy the sum rule in such QCD-like
  theories.
\end{abstract}

\maketitle

\section{Introduction}

The recent discovery of strongly coupled quark-gluon plasma at
RHIC \cite{Adcox:2004mh,Back:2004je,Arsene:2004fa,Adams:2005dq}
has spurred a significant theoretical effort towards understanding
the properties of matter described by strongly coupled quantum
field theories at finite temperature. The transport properties of
such theories are much more sensitive to the strength of the
coupling than stationary thermodynamic properties. In particular,
the near perfect fluidity in such theories is viewed as a
tell-tale sign of the strong coupling.

The transport properties of a theory are closely related to the
spectral functions of the operators such as stress-energy tensor.
In particular, viscosity can be determined from the low frequency
limit of the spectral function by the well-known Kubo formula. The
first-principles calculation of the spectral functions using
lattice Monte Carlo techniques is a challenging
task~\cite{Nakamura:2004sy,Aarts:2007wj,Meyer:2007ic,Meyer:2007dy,Huebner:2008as,Meyer:2008sn,Iqbal:2009xz,Meyer:2010ii},
especially in the low-frequency regime, and the prior knowledge of
the properties of spectral functions is essential. Therefore,
constraints on the spectral functions in the form of the sum rules
for the integral of the spectral function has been the subject of
recent
attention\cite{Kharzeev:2007wb,Karsch:2007jc,Romatschke:2009ng}.

The discovery of the AdS/CFT holographic
correspondence\cite{Maldacena:1997re,Gubser:1998bc,Witten:1998qj}
has opened new possibilities to study thermodynamics and transport
in strongly coupled theories (for reviews see
\cite{Aharony:1999ti,Erdmenger:2007cm,Son:2007vk,Myers:2008fv,Gubser:2009md,Ammon:2010zz,CasalderreySolana:2011us}).
Although not yet directly applicable to QCD, at least until we
know its holographic dual, these methods allow one to study
generic properties of the strongly coupled plasmas using model
theories, such as ${\cal
  N}=4$ SUSY YM, or holographic models which incorporate QCD-like
features such as confinement. In particular, the shear channel sum
rule derived in \cite{Romatschke:2009ng} has been verified in the
${\cal N}=4$ SUSY YM using AdS/CFT correspondence.

In this paper, we concentrate on the sum rule for the spectral
function in the bulk channel, corresponding to uniform dilatation
or isotropic expansion.  The bulk channel sum-rule is trivial in a
conformal theory such as ${\cal N}=4$ SUSY YM. However, nontrivial
sum rules in QCD do exists and have been a subject of recent
studies \cite{Kharzeev:2007wb,Karsch:2007jc,Romatschke:2009ng}.
Unlike the shear channel, the bulk channel correlation function
(and associated bulk viscosity) is sensitive not only to the
strength of the coupling, but also to the amount of the scaling
violation. Therefore we consider a holographic model with the
simplest mechanism of conformality violation, which is similar to
QCD. We shall assume that this model describes field theories
where scale invariance is broken by the presence of a scalar
operator ${\oO}$ with dimension $\Delta_+<4$ in the action. The
scale anomaly, $T^{\mu}_\mu$, is proportional to this operator,
and in QCD this role is played by the gluon condensate operator
$G_{\mu\nu}G^{\mu\nu}/\alpha_s$.

Thermodynamics and transport in such theories have been first
studied in Ref.~\cite{Gubser:2008ny,Gubser:2008yx,Gubser:2008sz}.
In particular, it has been shown that the speed of sound
$c_s=\sqrt{d\epsilon/dp}$ approaches the conformal value
$1/\sqrt3$ universally from {\em below}
\cite{Hohler:2009tv,Cherman:2009tw}.

The bulk sum rule has been also tested recently in Ref.~\cite{Springer:2010mw} in a theory whose holographic dual is a dilaton
gravity with the Chamblin-Reall dilaton
potential~\cite{Chamblin:1999ya}, which has the virtue of being
analytically tractable. However, the mechanism of the scaling violation in
such a putative field-theory (if the field-theory dual exists) would
differ from that in QCD. In particular, the speed of sound does not
approach its conformal value in the high-temperature limit.

In the general class of theories we consider, we shall find that the
temperature dependent part of the bulk spectral function $\deltaT{\rho(\omega)}$ obeys the following
sum rule
\begin{equation}\label{eq:sum-rule-Dplus}
  \left(3s\frac{\partial}{\partial s} - \Delta_+ \right)
\left(\epsilon-3p\right) = \frac{2}{\pi} \int_0^\infty
\frac{\deltaT {\rho\left(\omega\right)}}{\omega} d\omega.
\end{equation}
In the limit when the operator $\oO$ becomes marginal,
$\Delta_+\to4$, the sum rule coincides with the sum rule in QCD,
derived in \cite{Romatschke:2009ng}.

We shall also find that in this limit, which we shall refer to as
marginal, the sum rule in Eq.~(\ref{eq:sum-rule-Dplus}) exhibits the
same puzzle as discussed in
Refs.\cite{Moore:2008ws,CaronHuot:2009ns,Romatschke:2009ng} in the
context of QCD: the l.h.s. of the sum rule is of order $\alpha_s^3$, while
$\rho\sim \alpha_s^2$, where $\alpha_s$ is the QCD coupling. Here, in the marginal limit,
the l.h.s. of Eq.(1) is of order $\Delta_-^3$ while
$\rho\sim\Delta_-^2$, where
\begin{equation}
  \label{eq:d-minus}
  \Delta_-=
4-\Delta_+ .
\end{equation}
We shall demonstrate that a delicate cancellation indeed occurs
between the high frequency tail of the integral, $\omega \gg T$,
and the intermediate region of $\omega\sim T$, of the kind needed
to occur in QCD. Keeping only the leading ${\cal O}(\Delta_-^2)$
terms in $\deltaT{\rho}$, we find that the integral in
Eq.~(\ref{eq:sum-rule-Dplus}) converges. However, the integral
over the contribution of the subleading ${\cal O}(\Delta_-^3)$
terms in $\deltaT{\rho}$ has support over an interval of $\omega$
which stretches to infinity as
 $1/\Delta_-$ in the marginal limit
$\Delta_-\to0$. We evaluate the resulting additional ${\cal
  O}(\Delta_-^2)$
contribution from this long tail to the r.h.s. of
Eq.~(\ref{eq:sum-rule-Dplus}) analytically and show that it cancels
the ${\cal O}(\Delta_-^2)$ contribution from the $\omega\sim T$
region.

This paper is organized as follows. In the next section, we
present the definitions of quantities involved in the sum rule and
its derivation. In Section~\ref{sec:sum-rule}, we derive the sum
rule, Eq.~(\ref{eq:sum-rule-Dplus}), in the general class of
theories we consider. Section~\ref{sec:holographic-model}
introduces the holographic description of such theories.
Section~\ref{sec:two-point-functions} explains our method of
calculating the spectral function and the related Green's
functions. In Section~\ref{sec:large-freq-asympt}, we make
analytical calculation of the large $\omega$ asymptotics of
$\rho(\omega)$, which we use in Section~\ref{sec:spec} to subtract
from the numerically determined $\rho(\omega)$ to obtain
$\deltaT{\rho(\omega)}$. The sum rule is verified numerically in
Section~\ref{sec:sum} for a sample of values of $\Delta_-$. We
demonstrate the cancellation required to satisfy the sum rule in
the marginal limit in Section~\ref{sec:sum}. Cross-check of the
results with the existing analytical result for bulk viscosity
(Ref.\cite{Yarom:2009mw}) is made in Section~\ref{sec:viscosity}.
We conclude in Section~\ref{sec:conc}. Appendix~\ref{sec:appA}
contains the relevant results from~\cite{Hohler:2009tv}, used
throughout the paper.

\section{Definitions}
\label{sec:defs}

We shall consider only the response to homogeneous ($\bm q=0$)
perturbations and define the spectral function for the trace of the
stress-energy tensor $T^{\mu\nu}$, as usual, by
\begin{eqnarray}
\rho(\omega)& \equiv& - {\rm Im} \, G_R(\omega), \quad {\rm with}\\
G_R (\omega)& \equiv& - i \int_0^\infty dt \,  e^{i\omega t}\int d^3 \bm x
 \,\, \langle[T^\mu_\mu(x),T^\nu_\nu(0)]\rangle.
\end{eqnarray}
The definition of the retarded Green's function $G_R$ is subject to the
usual ambiguity due to the contribution of the product of the
operators at the same point at $x=0$. However, these contact terms do
not have imaginary parts and do not contribute to $\rho$.

It is convenient, as in \cite{Romatschke:2009ng}, to define Green's
functions of $T^{\mu\nu}$ by considering the response of the system to
the perturbation of the background metric
$g_{\mu\nu}=\eta_{\mu\nu}+\delta g_{\mu\nu}$ around the flat Minkowski metric
 $\eta_{\mu\nu}={\rm diag}(-1,1,1,1)$.  Euclidean time correlators
can be obtained similarly from variations of the Euclidean partition function
$Z[g]$:
\begin{equation}
  \label{eq:Z-g-Tmunu}
  \log Z[g]=\frac12 \int\! d^4x\,
\langle T^{\mu\nu}(x)\rangle \delta g_{\mu\nu}(x)
 +
\frac18\int\!\!\int\! d^4x\, d^4y\,
\langle T^{\mu\nu}(x)T^{\lambda\rho}(y)\rangle\,
\delta g_{\mu\nu}(x)\delta g_{\lambda\rho}(y)
+ {\cal O}(g^3).
\end{equation}
The correlation functions of the trace $\theta=T^\mu_\mu$ can be
also defined via variations of the partition function with respect to
 the metric variations of a special form (dilatations):
\begin{equation}
  \label{eq:g-omega}
  g(\Omega)_{\mu\nu}= \eta_{\mu\nu}e^{-2\Omega}
\end{equation}
\begin{equation}
  \label{eq:Z-Omega}
  \log Z[g(\Omega)]=-\int\!  d^4x\, \langle\theta(x)\rangle \Omega(x)
 +
\frac12\int\!\!\int\! d^4x\, d^4y\, \langle\theta(x)\theta(y)\rangle\,
\Omega(x)\Omega(y)
+ {\cal O}(\Omega^3).
\end{equation}
The two definitions of the two-point function of the trace differ by a
contact term:
\begin{eqnarray}
\langle \theta (x) \rangle &= &\eta_{\mu\nu}\langle T^{\mu\nu}(x)
\rangle\\
 \langle \theta(x)\theta(y) \rangle &=&
\eta_{\mu\nu}\eta_{\lambda\rho}\langle T^{\mu\nu}(x)
T^{\lambda\rho}(y)\rangle +2\delta^4(x-y)
\eta_{\mu\nu}\langle T^{\mu\nu}(x)\rangle
\label{eq:theta}
\end{eqnarray}

The corresponding retarded correlators can be
defined via linear response to perturbation $\Omega$:
\begin{equation}
  \label{eq:GR-theta}
  G_R(x,y) = \langle \theta(x)\theta(y) \rangle_R =
\frac{\partial}{\partial
\Omega(x)}\left(\sqrt{-g({\Omega(y)})} \langle\theta(y)\rangle\right)
\end{equation}
The definition (\ref{eq:GR-theta}) is convenient because the value
of its Fourier transform (for $\bm q=0$)
\begin{equation}
  \label{eq:theta-omega-def}
 G_R(\omega) = \langle \theta\theta \rangle_R(\omega)\equiv
 \int\! dt\, e^{i\omega t}\int d^3 \bm x \,\langle \theta(x)\theta(0) \rangle_R
\end{equation}
at vanishing frequency $\omega$ follows from the conservation of
entropy in the ideal hydrodynamics
\cite{Romatschke:2009ng}:
\begin{equation}
  \label{eq:theta-theta-0-ds}
   \langle \theta\theta \rangle_R(0)
=\frac{\partial}{\partial
\Omega}\left(\sqrt{-g({\Omega})} \langle\theta\rangle\right)
= \left( 3 s \frac{\partial}{\partial s} -4\right) \langle\theta\rangle
\end{equation}
where $s$ is the entropy density.

\section{The sum rule}
\label{sec:sum-rule}

The sum rule derived in  \cite{Romatschke:2009ng} applies to the zero
temperature subtracted Green's function
\begin{equation}
  \label{eq:delta-GR}
  \deltaT {G_R} \equiv G_R - G_R^{(T=0)}
\end{equation}
and its imaginary part on the real axis $\deltaT \rho=-{\rm Im}\,\deltaT {G_R}$:
\begin{equation}
  \label{eq:sum-rule}
  \deltaT {G_R(i\infty)} -\deltaT {G_R(0)}
=\frac2\pi \int_0^\infty \!  d\omega\, \frac{\deltaT
{\rho\left(\omega\right)}}{\omega}.
\end{equation}
Using Eq.~(\ref{eq:theta-theta-0-ds}) and  $\deltaT
{\langle\theta\rangle} = 3p-\epsilon$, one can write
\begin{equation}\label{eq:delta-T-GR-0}
  \deltaT {G_R(0)} = - \left( 3 s \frac{\partial}{\partial s} -4\right)(\epsilon-3p)
\end{equation}
where $\epsilon=\deltaT{\langle T^{00}\rangle}$ and $p=\deltaT{\langle T^{11}\rangle}$ are equilibrium thermal energy and pressure
at given entropy density $s$.  In QCD, due to the asymptotic freedom,
$\deltaT {G_R(i\infty)}=0$ and one obtains the sum rule found in
\cite{Romatschke:2009ng}:
\begin{equation}
  \label{eq:sum-rule-e-3p}
\left( 3 s \frac{\partial}{\partial s} -4\right)(\epsilon-3p)
=
\frac2\pi \int \!  d\omega\,
\frac{\deltaT {\rho\left(\omega\right)}}{\omega}
\end{equation}

In this paper, we consider a generic conformal field theory
perturbed by an operator $\oO$ of dimension $\Delta_+$ sourced by
the field $c$. We place this theory on a nontrivial gravitational
background given by metric $g_{\mu\nu}$. The change of the
partition function of the theory under dilatations
(\ref{eq:g-omega}) is equivalent to the rescaling of the only
dimensionful external field $c$, the source of the operator $\oO$:
\begin{equation}
  \label{eq:Zc}
  Z[g(\Omega),c] = Z[g(0),e^{-{\Delta_-}\Omega}c],
\end{equation}
where $\Delta_-=4-\Delta_+$. Thus
\begin{equation}
  \label{eq:theta-c}
  \frac{\delta\log Z}{\delta\Omega}
=-\Delta_- e^{-{\Delta_-}\Omega}c \frac{\delta\log Z}{\delta c}
=\Delta_- e^{-{\Delta_-}\Omega} c\langle\oO\rangle.
\end{equation}
Thus
\begin{equation}
  \label{eq:theta-O}
   \langle\theta(x)\rangle =  -\Delta_- c\langle\oO(x)\rangle,
\end{equation}
and
\begin{equation}
  \label{eq:theta-theta-O}
   \langle\theta(x)\theta(y)\rangle =  \Delta_-^2
   c^2\langle\oO(x)\oO(y)\rangle
- \Delta_-^2 c\langle \oO\rangle\delta^4(x-y).
\end{equation}
To evaluate the Fourier transform of the correlation
function~(\ref{eq:theta-theta-O}) at large (imaginary) frequency
$\omega$, we can use the operator product expansion (OPE) for
$\oO(x)\oO(y)$. The leading contribution comes from the unit
operator, and behaves as $\omega^{2\Delta_+-4}$, but cancels in
$\deltaT {G_R}$ because it is independent of temperature.
The contribution of an operator of dimension $\Delta$  comes with
the Wilson coefficient which behaves as
$\omega^{2\Delta_+-4-\Delta}$. Assuming that there are no
operators (with vacuum quantum numbers) of dimension equal to or
lower than $2\Delta_+-4$, we conclude that the contribution of the
first term in Eq.~(\ref{eq:theta-theta-O}) vanishes in the
$\omega\to i\infty$ limit. The assumption implies, in particular,
that $\Delta_+<4$, i.e., $\Delta_->0$. The contribution of the
last term in Eq.~(\ref{eq:theta-theta-O}) then gives the value for
$G_R(i\infty)$:
\begin{equation}
  \label{eq:GR-infty}
  \deltaT{G_R(i\infty)}
= -\deltaT{\langle\theta\theta\rangle(i\infty)}
= \Delta_-^2\, c\ \deltaT {\langle\oO\rangle}
= \Delta_-\,(\epsilon-3p),
\end{equation}
where we used the fact that analytical continuation of a retarded
correlation function to Matsubara frequencies on the imaginary
axis equals negative of the Euclidean correlator.  We shall verify
Eq.~(\ref{eq:GR-infty}) explicitly and analytically in the
holographic model.  Combining
Eqs.~(\ref{eq:delta-T-GR-0}),~(\ref{eq:GR-infty})
and~(\ref{eq:sum-rule}) we obtain Eq.~(\ref{eq:sum-rule-Dplus}).

\section{Holographic model}
\label{sec:holographic-model}

As discussed above, we will consider a four-dimensional (4D)
conformal theory in which the conformal symmetry is broken by the
operator $\oO$ with scaling dimension $\Delta_+$. The
holographically dual description of such a theory must therefore
include a five-dimensional (5D) scalar field with mass
$m_5^2=-\Delta_+\Delta_-$.  As a model for the QCD thermodynamics,
such a theory was first considered in
\cite{Gubser:2008ny,Gubser:2008yx}. It models in the most
straightforward way the breaking of the scaling invariance in QCD.
The minimum set of fields required to describe correlators of the
stress energy tensor and of the operator $\oO$ include the 5D
metric $g_{MN}$ ($M,N=0,1,2,3,z)$ and the scalar field (dilaton)
$\phi$. The minimal action is thus given by
\begin{equation}
  \label{eq:S5}
  \begin{split}
S_5 &= S_\bulk + S_{\rm GH} \\
 &= \frac{1}{2 \kappa^2} \left[\int_M \!\! d^5x \; \sqrt{-g}\,
\left(R - V\left(\phi\right) - \frac{1}{2} \left(\partial
\phi\right)^2 \right) - 2 \int_{\partial M} \!\!\!\! d^4x \;
\sqrt{-\gamma}\, K \right],
\end{split}
\end{equation}
where $R$ is the
Ricci scalar, $g$ is the determinant of the metric, $\gamma$ is
the determinant of the induced metric on the UV boundary $\partial
M$ at $z=0$, $K$ is the extrinsic curvature on $\partial M$, and $\kappa^2$
is the 5D Einstein gravitational constant. The value of $\kappa^2$
is inversely proportional to the number of the degrees of freedom
in the dual four-dimensional theory, e.g., $N_c^2$ in a gauge
theory with large number of colors $N_c$. The smallness of
$\kappa^2$ (i.e., the largeness of the number of colors) controls
the semiclassical approximation which we use. The last term in the
action is the Gibbons-Hawking term, which removes the boundary
terms arising upon integration by parts of the terms in $R$ linear
in second derivatives of the metric~\cite{Gibbons:1976ue}.

The potential for the dilaton, $V(\phi)$, is the function which
was tuned in~\cite{Gubser:2008ny,Gubser:2008yx} to best ``mimic''
the QCD equation of state. Here, similarly to
\cite{Hohler:2009tv}, we shall concentrate on the results which
are universal in the class of models described by
Eq.~(\ref{eq:S5}) with any (sensible) potential. For example,
Ref.~\cite{Hohler:2009tv} found that the speed of sound approaches
the conformal value $1/\sqrt3$ as $T\to\infty$ universally from
below in such theories. Here we shall also make use of the large
$T$ limit to the extent that it makes only the curvature of the
potential $V''(\phi)=m_5^2=\Delta_+ \left(\Delta_+-4\right)$
matter.

The negative cosmological constant provided by choosing $V(0)=-12$
ensures necessary asymptotics of the metric near the boundary
$g_{MN}\sim z^{-2}$ while the dilaton field asymptotics is fixed by
$V''(\phi)$: $\phi\sim z^{\Delta_-}$. The singular behavior at $z=0$
can be regulated by setting the boundary conditions at $z=\varepsilon$
and taking the UV regulator $\varepsilon$ to 0 after necessary
renormalization. The rules of the holographic correspondence require
us to extremize the action $S_5$ w.r.t. the metric and the dilaton field
subject to the boundary conditions on the UV boundary
$z=\varepsilon\to0$:
\begin{equation}
  \label{eq:b-c}
  g_{\mu\nu}(x,z)\big|_{z=\varepsilon}=g_{\mu\nu}(x)\,\varepsilon^{-2},
\qquad
\phi(x,z)\big|_{z=\varepsilon} = c(x)\,\varepsilon^{\Delta_-}\,.
\end{equation}
The holographic duality then implies that the extremal value of the
Euclidean 5D action as a functional of the boundary values $g_{\mu\nu}(x)$ and
$c(x)$ is the same functional as $-\log Z[g_{\mu\nu}, c]$ in the 4D
theory. In order to calculate the Green's function $\langle\theta\theta\rangle$ we use
Eq.~(\ref{eq:theta}) and the relationship (\ref{eq:Z-g-Tmunu})
between the derivatives of $\log Z$ and the correlators of $T^{\mu\nu}$.

The one-point  functions $\langle T^{\mu\nu}\rangle$ are
determined by the solutions of the equations of motion with
boundary conditions homogeneous in the 4D coordinates $x^\lambda$.
We follow Ref.~\cite{Hohler:2009tv} to determine the solution to
the equations of motion homogeneous in 4D. In the gauge chosen in
Ref.~\cite{Hohler:2009tv}, the metric has the form
\begin{equation} \label{eq:metric}
ds^2 = \frac{1}{z^2} \left( -f\left(z\right) dt^2 + d\bm{x}^2 +
e^{2 B\left(z\right)} \frac{dz^2}{f\left(z\right)}\right)
\equiv g^{(0)}_{MN} dx^M d x^N,
\end{equation}
where $f(z)$ and $B(z)$ are functions only of the extra
dimensional coordinate $z$, and $f(z)$ has a simple zero at some value of
$z=z_H$. The functions, $f$ and $B$, and
the background dilaton field, $\phi$, are all determined by extremizing the
action. The details can be found in \cite{Hohler:2009tv}
and for completeness presented in Appendix \ref{sec:appA}.

The extremum of the action with boundary
conditions~(\ref{eq:b-c}) is a
one-parameter family of solutions to the equations of motion. A
convenient choice of the parameter is the enthalpy density, $w$, which
arises as an integration constant (see Eq.~(\ref{eq:ein2b})). The temperature is
determined, as usual, by considering the periodicity in the Euclidean
time necessary to avoid conical singularity at $z=z_H$:
\begin{equation}
T = \frac{1}{4\pi} e^{-B(z_H)}|f'(z_H)|,
\end{equation}
and is related to $w$ via equation of state (see e.g.
Eq.~(\ref{eq:w-T})).

\section{Two-point functions}
\label{sec:two-point-functions}

The calculation of two-point correlation functions requires us to
consider non-homogeneous solutions to the equations of motion. We
only need to consider infinitesimal variations around the
homogeneous solution (\ref{eq:metric}) and expand the action $S_5$
to quadratic order in these variations. We parameterize these
inhomogeneous solutions as
\begin{equation}
  \label{eq:g-h}
  g_{MN} = g^{(0)}_{MN}(1+ h_{MN})
\end{equation}
where summation over $M,N$ is not implied. We exercise the freedom
of the gauge choice to set $\phi(x,z)=\phi^{(0)}(z)$, i.e., we set
variations of $\phi$ around homogeneous solution to 0 (we can do
this as long as we are not interested in calculating correlation
functions of $\oO$, i.e., as long as  $c$ remains homogeneous). We
find that this gauge choice provides nontrivial simplification of
the equations of motion, allowing us to reduce them to a second
order equation for metric variation $H$, instead of the third.

We then consider solution homogeneous in the spatial coordinates
$x^i$, and depending only on $x^0\equiv t$ and $z$, since we are
interested in $\bm q=0$ variations of $g_{\mu\nu}$. For such
solutions we can use the remaining gauge freedom to set $h_{z\mu}=0$,
but not $h_{zz}$.

We can also use $O(3)$ symmetry
to simplify our analysis by separating the spatial part of metric
perturbations into trace and traceless parts:
\begin{equation}
  \label{eq:h-H-ht}
  h_{ij} =  H\delta_{ij}+ h^T_{ij}.
\end{equation}
At quadratic order,  $H$ mixes with components $h_{00}$ and
 $h_{zz}$, but decouples from the traceless part $h^{T}_{ij}$ as
well as from the off-diagonal components $h_{0i}$. Therefore, we shall
focus only on the metric perturbations $H$, $h_{00}$ and
$h_{zz}$. We can express the coupling of the metric perturbation
to $T^{\mu\nu}$ as
\begin{equation}
  \label{eq:h-T}
  h_{\mu\nu} T^{\mu\nu} = H \Sigma + h^T_{ij}T_T^{ij}
+ 2h_{0i}T^{0i}+ h_{00}T^{00},
\end{equation}
where $\Sigma=T^{ij}\delta_{ij}$ is the trace and $T_T^{ij}=T^{ij} -
\frac13\Sigma\delta^{ij}$ is the traceless part of the {\em stress\/}
tensor $T^{ij}$. The two-point function $\langle\theta\theta\rangle$
in Eq.~(\ref{eq:theta}), since
$\eta_{\mu\nu}T^{\mu\nu} = \Sigma - T^{00}$, can be expressed in terms
of correlation functions of $T^{00}$ and $\Sigma$ using
\begin{equation}
\eta_{\mu\nu}\eta_{\lambda\rho}\langle T^{\mu\nu}(x)
T^{\lambda\rho}(y)\rangle = \langle
T^{00}(x)T^{00}(y) \rangle - 2 \langle
T^{00}(x)\Sigma(y) \rangle +  \langle
\Sigma(x)\Sigma(y) \rangle.
\end{equation}
The corresponding retarded correlators can be found using
holographic correspondence and the recipe~\cite{Son:2002sd}:
\begin{align}
\langle T^{00}(t)T^{00}(t')\rangle_R & =-4
\left.\frac{\delta^2 S_5}{\delta
h_{00}(t,z)\delta h_{00}(t',z)}\right|_{z=\varepsilon},\label{eq:T0000}\\
\langle T^{00}(t)\Sigma(t')\rangle_R & =-4
\left.\frac{\delta^2 S_5}{\delta
h_{00}(t,z)\delta H(t',z)}\right|_{z=\varepsilon},\label{eq:T00ss}\\
\langle \Sigma(t)\Sigma(t') \rangle_R & =-4
\left.\frac{\delta^2 S_5}{\delta H(t,z)\delta
H(t',z)}\right|_{z=\varepsilon}.\label{eq:Tssss}
\end{align}
Since we need the Fourier transform of $\langle\theta\theta\rangle$ it
would be convenient to express $S_5$ directly in terms of the Fourier
modes of $h_{\mu\nu}$ defined as
\begin{equation}\label{eq:h-omega}
  h_{MN}(t,z) = \int \frac{d\omega }{2\pi}
h_{MN}(\omega,z) e^{-i\omega t}.
\end{equation}

The metric perturbations $H$,
$h_{zz}$ and $h_{00}$, defined by Eqs.~(\ref{eq:g-h}) and~(\ref{eq:h-H-ht}),
satisfy a set of equations derived from
the linearized Einstein's equations:
\begin{eqnarray} \label{eq:eom}
H''(\omega,z) &=& H'(\omega,z) \left(\frac{2}{z} + B'(z) - \frac{f'(z)}{f(z)} - \frac{B''(z)}{B'(z)}\right) +H(\omega,z) \left(-\frac{\omega^2 e^{2 B(z)}}{f(z)^2} + \frac{f'(z)}{2 z f(z)}+\frac{f'(z)}{2 f(z)}\frac{B''(z)}{B'(z)} \right),\\
h_{00}'(\omega,z) &=& H'(\omega,z) \left(-1-z
B'(z)+\frac{z}{2}\frac{f'(z)}{f(z)}\right) + H(\omega,z)
\left(2\frac{f'(z)}{f(z)}+\frac{z}{2}\frac{f'(z)B'(z)}{f(z)^2} -
\omega^2  \frac{z\,e^{2 B(z)}}{f(z)^2} -
\frac{z}{2}\frac{f'(z)^2}{f(z)^2}\right),\\\label{eq:eom-hzz}
h_{zz}(\omega,z) &=& -z\,H'(\omega,z)  +H(\omega,z) \left(\frac{z}{2}\frac{f'(z)}{f(z)}\right),
\end{eqnarray}
with the primes denoting derivatives with respect to $z$. In order
for Eqs.~(\ref{eq:T0000})--(\ref{eq:Tssss}) to give the {\em
retarded} correlation functions, the solutions to
Eqs.~(\ref{eq:eom})--(\ref{eq:eom-hzz})
 must satisfy the in-falling wave
conditions at the horizon $z=z_H$.

In order to calculate the two-point functions, we expand the
holographic action in Eq.~(\ref{eq:S5}) to quadratic order in metric
variations around the homogeneous background. This expansion can be
written in a compact form; the bulk part of the action is given
by,
\begin{equation}
S_\bulk =\frac{1}{2 \kappa^2} \int \! \frac{d\omega}{2 \pi}
\, d^3x \,dz \,  \frac{e^{-B(z)}f(z)}{z^3} \left({\bm
H}^{\prime\prime \dagger} M_1 {\bm H} + {\bm H}^{\prime\dagger} M_2 {\bm
H}^\prime + {\bm H}^{\prime \dagger} M_3 {\bm H}
 + {\bm H}^{\dagger} M_4 {\bm H} + ({\rm c.c.})\right),
\end{equation}
where the dagger refers to the transposed complex conjugate ($\bm
H^*(\omega,z)=\bm H(-\omega,z)$), and
\begin{equation}
\begin{split}
& {\bm H}=\left(\begin{array}{c}
H(\omega,z)\\h_{00}(\omega,z)\\h_{zz}(\omega,z)
\end{array}\right); \qquad  M_1 = \frac{1}{2}\left(
\begin{array}{ccc} -3&3&3\\3&1&-1\\0&0&0
\end{array} \right);  \qquad M_2 = \frac{1}{4}  \left(\begin{array}{ccc} 0&3&3\\3&2&-1\\3&-1&0\end{array} \right);\\
& M_3 = \frac{2}{z} \left(\begin{array}{ccc}
3&-3&-3\\-3&-1&1\\-3&1&3\end{array}\right) + \frac{1}{2}
 B'(z)\left(\begin{array}{ccc}
3&-3&-3\\-3&-1&1\\0&0&0
\end{array}\right)  + \frac{1}{4} \frac{f'(z)}{f(z)}
\left( \begin{array}{ccc} -6&6&6\\9&3&-3\\3&-1&-3
\end{array} \right);\\
& M_4 = \frac{1}{4z} \left(\frac{4}{z}+
B'(z)-\frac{f'(z)}{f(z)}\right) \left(\begin{array}{ccc}
-3&3&12\\3&1&-4\\12&-4&-9
\end{array} \right) - \frac{3}{4}\frac{\omega^2 e^{2 B(z)}}{f(z)^2}
\left(\begin{array}{ccc} 2&0&1\\0&0&0\\1&0&0
\end{array} \right).
\end{split}
\end{equation}
Integrating by parts and using equation of motion, we reduce the
action at the extremum to the boundary term:
\begin{multline}
S_\bulk=-\frac{1}{2 \kappa^2} \int \! \frac{d\omega}{2 \pi} \,
d^3x \,  \frac{e^{-B(z)}f(z)}{z^3} \Bigg({\bm H}^{\prime\dagger}
M_1 {\bm H} - {\bm H}^{\dagger} M_1 {\bm H}
\left(\frac{f'(z)}{f(z)} -B'(z)-\frac{3}{z}\right) - {\bm
H}^{\dagger} M_1 {\bm H}^{\prime} \\ + 2 {\bm H}^{\dagger} M_2
{\bm H}^\prime + {\bm H}^{\dagger} M_3 {\bm H}
\Bigg)\Bigg|_{z=\varepsilon}.
\end{multline}
The action also receives a
contribution from the Gibbons-Hawking boundary term:
\begin{equation}
\begin{split}
S_{\rm GH} =  \frac{1}{2 \kappa^2}\int \! \frac{d\omega}{2 \pi} \,
d^3x \,   \frac{e^{-B(z)}f(z)}{z^3} \Big({\bm H}^{ \prime \dagger}
M_5 {\bm H} +
 {\bm H}^{\dagger} M_6 {\bm H} + ({\rm c.c.})\Big)\Big|_{z=\epsilon}; \quad {\rm with}\\
M_5 = M_1 \qquad {\rm and} \qquad M_6 = \frac{1}{8}
\left(\frac{8}{z}-\frac{f'(z)}{f(z)}\right) \left(
\begin{array}{ccc}
3&-3&-3\\-3&-1&1\\-3&1&3
\end{array} \right).
\end{split}
\end{equation}
Combining all of this, the action at the extremum can be expressed in
terms of the boundary values of $h_{00}$, $H$ and $H'$:
\begin{multline}
S_5 = \frac{1}{2 \kappa^2}\int \! \frac{d\omega}{2 \pi} \, d^3x
\frac{e^{-B(z)}f(z)}{z^3}
\,\left(-\frac{3}{2z} 
h_{00}^*(\omega,z)h_{00}(\omega,z) + \left(-
\frac{9}{2z}+\frac{3}{4} \frac{f'(z)}{f(z)}\right)
\big( h_{00}^*(\omega,z) H(\omega,z)+
({\rm c.c.})
\big)\right. \\
+\left(\frac{9}{2z^2}+\frac{3}{4}\frac{B'(z)f'(z)}{f(z)}-\frac{3}{8}\frac{f'(z)^2}{f(z)^2}-\frac{3}{2}
\frac{\omega^2 e^{2 B(z)}}{f(z)^2} \right)
z 
\, H^*(\omega,z) H(\omega,z) \\
- \left.\left.\frac{3}{2}
z B'(z)\, H^*(\omega,z)H^\prime(\omega,z)\right)
\right|_{z=\varepsilon}.
\end{multline}
From this one can immediately read off the expressions for the Fourier
transforms of the stress-energy two-point functions:
\begin{align}
\langle T^{00}T^{00} \rangle(\omega) &=  \left.\frac{1}{2
\kappa^2}\frac{6
e^{-B(z)}f(z)}{z^4}  \right|_{z=\varepsilon}
=-\langle T^{00}\rangle ;\label{eq:0000}\\
\langle T^{00} \Sigma \rangle(\omega) & =  \left.\frac{3}{2
\kappa^2}\left(\frac{6 e^{-B(z)} f(z)}{z^4} - \frac{e^{-B(z)}
f'(z)}{z^3}\right)
\right|_{z=\varepsilon}
=\langle \Sigma \rangle\label{eq:00xx}; \\
\langle \Sigma\Sigma \rangle(\omega)  & =  \left.
\frac{9}{2\kappa^2}\frac{e^{-B(z)} f(z)}{z^2} \left(
-\frac{2}{z^2} - \frac{B'(z) f'(z)}{3 f(z)} + \frac{ f'(z)^2}{6
f(z)^2} + \frac{2 \omega^2 e^{2 B(z)}}{3f(z)^2} + \frac{2
B'(z)}{3} 
\frac{H^\prime(\omega,z)}{H(\omega,z)}\right)\right|_{z=\varepsilon},
\label{eq:xxxx}
\end{align}
where in the last equality in Eqs.~(\ref{eq:0000})
and~(\ref{eq:00xx}) we used expressions for one-point functions
Ref.~\cite{Hohler:2009tv}. As expected, the correlators involving
$T^{00}$ are frequency independent and satisfy energy-momentum
conservation Ward identies~Ref.~\cite{Policastro:2002tn}.
 The above expressions contain  contributions divergent in the limit
$\varepsilon \rightarrow 0$. They can be removed by subtracting a polynomial of $\omega^2$ sufficient
to cancel the divergences.
These terms do not affect $\rho(\omega)$.

Putting this all together, the finite (as $\varepsilon\to0$) part of the bulk Green's function is
found to be,
\begin{equation} \label{eq:tt}
G_R(\omega)-P(\omega^2) = 4 (\epsilon -3 p) + \frac{6}{2\kappa^2}
\left.\frac{e^{-B(z)} f(z) B'(z)}{z^2}
\frac{H^\prime(\omega,z)}{H(\omega,z)}\right|_{z=\varepsilon},
\end{equation}
where the polynomial $P(\omega^2)= 4\langle
T^{00}-\Sigma\rangle_{T=0} +\omega^2\cdot
3e^{B(\varepsilon)}/(\kappa^2\varepsilon^2)$ combines the UV
divergent, temperature independent contact terms. The bulk
spectral function is then given by
\begin{equation} \label{eq:bulk}
\rho (\omega) = - \frac{6}{2\kappa^2} \left.
\frac{e^{-B(z)} f(z) B'(z)}{z^2} \, {\rm Im} \frac{H^\prime(\omega,z)}{H(\omega,z)}\right|_{z=\varepsilon}.
\end{equation}


\section{Large frequency asymptotics}
\label{sec:large-freq-asympt}

In order to calculate the bulk retarded Green's function and the bulk
spectral function, one needs to solve the equation of motion for the
field $H(\omega,z)$, Eq.~(\ref{eq:eom}).  In general, this cannot be done
analytically. However, in the large $\omega$ limit, the solution can
be found similarly to the Born approximation in quantum mechanics.

We perform a Louiville transformation to bring
Eq.~(\ref{eq:eom}) to the Schr\"{o}dinger form.
In terms of the new coordinate $x$, given by
\begin{equation}
x = \int^z \frac{e^{B(z')}}{f(z')} dz',
\end{equation}
and the new function $\Psi(x)$, given by
\begin{equation} \label{eq:htophi}
H(z) = z B'(z)^{-1/2} \Psi(x),
\end{equation}
Eq.~(\ref{eq:eom}) takes the form
\begin{equation} \label{eq:eomphi}
\frac{d^2{\Psi}(x)}{dx^2}+ \Psi(x) \left(\omega^2 - U_{\rm Sch}(x)\right)
=0,
\end{equation}
with the
Schr\"{o}dinger potential, as an implicit function of $x$, is given
by
\begin{equation} \label{eq:psi}
U_{\rm Sch}(x) =
\frac{ f(z)^2}{ e^{2 B(z)}} \left(\frac{2}{z^2} -
\frac{1}{z}\frac{B''(z)}{B'(z)}
-\frac{1}{4}\frac{B''(z)^2}{B'(z)^2}
+\frac{1}{2}\frac{B'''(z)}{B'(z)} -\frac{1}{2z}\frac{f'(z)}{f(z)}
+\frac{f'(z)}{f(z)}\frac{B''(z)}{B'(z)}+\frac{1}{z}
B'(z)-\frac{1}{2} B''(z)\right).
\end{equation}
The in-falling boundary condition on $H(z)$ at the horizon $z=z_H$
transforms into the outgoing wave condition on $\Psi(x)$ at
$x=\infty$. The bulk spectral function can
be expressed in terms of the new function $\Psi$ as
\begin{equation}\label{eq:rho-Psi}
\rho = -\frac{1}{2\kappa^2}
\left.
\frac{6B'(z)}{z^2}\,
{\rm Im} \frac{
\Psi^\prime(x)}{\Psi(x)}
\right|_{z=\varepsilon}\,,
\end{equation}
or somewhat more intuitively (using Eq.~(\ref{eq:ein1})) as
\begin{equation}\label{eq:rho-Psi2}
\rho =
\left.
\frac{\phi'(z)^2}{2\kappa^2z}\,
\frac{
{\rm Im} \left[
\Psi^*(x) \Psi^\prime(x)\right]
}{|\Psi(x)|^2}
\right|_{z=\varepsilon}
\end{equation}
which has the quantum mechanical interpretation of the probability
flux for the wave function $\Psi$ (normalized as
$|\Psi(x)|^2=\phi'(\varepsilon)^2/(2\kappa^2\varepsilon)$
at $z=\varepsilon$). The
positivity of the spectral function can be seen then as a consequence of
the conservation of the flux and the outgoing wave boundary condition
at $x=\infty$.

At large $\omega$, we would like to treat the potential as a
perturbation and calculate the ``wave function'' $\Psi$ using the
Born approximation. There is one difficulty, however: the
Schr\"{o}dinger potential $U_{\rm Sch}(x)$ diverges at $x=0$. We
shall separate the leading divergence explicitly
\begin{equation} \label{eq:phi2}
\frac{d^2{\Psi}(x)}{dx^2}+ \Psi(x) \left(\omega^2 -
  \frac{\left(3 - 2\Delta_-\right)\left(2\Delta_+-3\right)}{4x^2}
-  \delta U_{\rm Sch}(x)\right)
=0
\end{equation}
and treat the remaining part of the potential,  $\delta U_{\rm
  Sch}(x)$, as a perturbation. We can calculate $\Psi(x)$ iteratively,
using the Green's function  method. The solution to
Eq.~(\ref{eq:phi2}) with $\delta U_{\rm Sch}=0$, satisfying the
outgoing wave boundary condition at $x=\infty$, is given by, up to
unimportant normalization,
\begin{equation} \label{eq:phi0sol}
\Psi_0(x) =  (\omega x)^{1/2} H_{\nu}^{(1)}\!
\left(\omega x\right)
\end{equation}
where $\nu = 2-\Delta_-$, $H_\nu^{(1)}$ is the
Hankel function the first kind. Substituting $\Psi_0$ into
(\ref{eq:htophi}) and then into (\ref{eq:tt}) we find, using
Eq.~(\ref{eq:b-small-z}) for $B'(z)$,
\begin{equation} \label{eq:grlw}
G_R(\omega) - P(\omega^2)= \frac{c^2}{2\kappa^2}
\frac{ 2\pi \Delta_-^2}{\Gamma(2-\Delta_-)^2}\,
\left(\cot(\pi \Delta_-)-i\right) \,
\left(\frac{\omega}{2}\right)^{\Delta_+-\Delta_-}
 - \frac{c\, d}{2\kappa^2}\, \Delta_-^2
\left(\Delta_+-\Delta_-\right)
+ \ldots
\end{equation}
where the dots stand for corrections to the leading order which would come from
iterations of $\delta U_{\rm Sch}$.
The first term on the r.h.s. is the leading term in
the $\omega\to\infty$ limit. It is temperature independent and has
correct $\omega$-scaling to be identified with the leading contribution of the
unit operator to the OPE of $\langle\oO\oO\rangle$ in
Eq.~(\ref{eq:theta-theta-O}).
This term has nontrivial, but temperature independent, imaginary part
and it is subtracted when we calculate  $\deltaT {G_R}$.

The
second term on the r.h.s. of Eq.~(\ref{eq:grlw}) gives
\begin{equation}
\deltaT{G_R (i\infty)} = - \frac{c\, \deltaT d}{2\kappa^2}\, \Delta_-^2\left(\Delta_+-\Delta_-\right) = \Delta_- (\epsilon - 3p).
\end{equation}
where we used Eq.~(\ref{eq:e-p}). This agrees with
 Eq.~(\ref{eq:GR-infty}) as expected.

The leading correction, $\Psi_1$, to $\Psi_0$ can be found, similarly to the Born
approximation, by using  Green's function $G(x,x')$:
\begin{equation}\label{eq:psi1-Gpsi0}
\Psi_1(x) = \int dx' G(x,x') \Psi_0(x') \delta U_{\rm Sch}(x').
\end{equation}
The Green's function satisfies equation
\begin{equation}
\frac{d^2 G(x,x')}{d x^2} + G(x,x')
\left(\omega^2-\frac{\left(2\Delta_--3\right)\left(2\Delta_+-3\right)}{4x^2}\right)
= \delta(x-x'),
\end{equation}
with boundary condition $G(x,x')=0$ at $x=\varepsilon$ and outgoing
wave condition at $x=\infty$. The solution is given by
\begin{equation}
G(x,x') = \left\{\begin{array}{ll} \gamma_<(x') \,(\omega
x)^{1/2}
J_\nu(\omega x) & x<x'\\
\gamma_>(x')\, (\omega x)^{1/2} H_\nu^{(1)}(\omega x) & x>x'
\end{array}\right.
\end{equation}
where
\begin{equation}
\gamma_<(x')=-\frac{i\pi}{2\omega}\,(\omega x')^{1/2}
H_\nu^{(1)}\!\!\left(\omega x'\right) \quad {\rm and} \quad
\gamma_>(x') =-\frac{i\pi}{2\omega}\,(\omega x')^{1/2}
J_\nu\!\left(\omega x'\right).
\end{equation}
Substituting into Eq.~(\ref{eq:psi1-Gpsi0}) we find
\begin{equation}\label{eq:psi1sol}
\Psi_1(x) = -\frac{i \pi}{2} (\omega x)^{1/2}\left[J_\nu(\omega x)
\int_x^\infty \! x' \delta U_{\rm Sch}(x')
\left(H_\nu^{(1)}(\omega x')\right)^2 dx'+ H_\nu^{(1)}(\omega x)
\int_\varepsilon^x x' \delta U_{\rm Sch}(x') H_\nu^{(1)}(\omega
x') J_\nu(\omega x') dx'\right].
\end{equation}

To zeroth order in $\delta U_{\rm Sch}$, $\Psi=\Psi_0$ and Eq.~(\ref{eq:rho-Psi})
gives:
\begin{eqnarray}
\rho_0(\omega) &=& \frac{c^2}{2\kappa^2}
\frac{2\pi\Delta_-^2}{\Gamma(2-\Delta_-)^2}
\left(\frac{\omega}{2}\right)^{4-2\Delta_-}.\label{eq:rho0}
\end{eqnarray}
To next order in $\delta U_{\rm Sch}$, $\Psi=\Psi_0+\Psi_1$ and
$\rho=\rho_0+\rho_1$, where
\begin{eqnarray}
\rho_1(\omega) &=& \pi\rho_0(\omega)\int_0^\infty x\, \delta U_{\rm Sch}(x) J_\nu(\omega x) Y_\nu(\omega x) dx\,.
\label{eq:rhoint}
\end{eqnarray}

In the limit $\omega\to\infty$, the integral in
Eq.~(\ref{eq:rhoint}) is dominated by the region of small
$x\sim\omega^{-1}$. Expanding $\delta U_{\rm Sch}$ in powers of
$x$, we generate a $1/\omega$ series (possibly asymptotic) for the
integral.

To enable further analytic calculation, we shall limit it to
leading terms in the expansion in powers of $c^2$. As we also
discuss in Appendix~\ref{sec:appA}, this corresponds to the regime
of high temperatures, i.e., $T\gg c^{1/\Delta_-}$, since $c$ is
the only dimensionful parameter in the theory. In addition to
enabling analytic calculations, this limit has the virtue of
yielding results which do not depend on the form of the dilaton
potential beyond its curvature at the minimum (related to
$\Delta_-$).  In this regime, to the order we work, we can set $c$
to zero in $U_{\rm Sch}$.  The leading term in the Taylor
expansion of $U_{\rm Sch}$ is given by
\begin{equation}
 \delta U_{\rm Sch}(x)= \frac{3}{20} \Delta_-\Delta_+  \bar w\, x^2 +
 \cO(x^6),
\end{equation}
where $\bar w=2\kappa^2 w$, as defined in Appendix~\ref{sec:appA}, and
\begin{equation}
  \label{eq:rho1-rho0}
 {\rho_1}=-\frac{1}{20} \frac{\bar w}{\omega^4} \Delta_-\Delta_+
(1-\Delta_-)(\Delta_+-1)(\Delta_+-\Delta_-){\rho_0}(\omega).
\end{equation}
Combining Eq.~(\ref{eq:rho0}) and Eq.~(\ref{eq:rho1-rho0}) we see that
$\rho_1\sim \omega^{-2\Delta_-}\to0$ as $\omega\to\infty$.

We note that, strictly speaking, not all $\cO(c^2)$ corrections to
this result are negligible, since some of them grow with $\omega$.
 However, as one can check, those corrections do
not depend on $w$ (i.e., on $T$) and are part of the $\cO(c^2\omega^{-2\Delta_-})$
corrections to the Wilson coefficient of the unit operator in the
OPE of $\langle\oO\oO\rangle$. These terms would be also
subtracted if $\deltaT{\rho(\omega)}$ was calculated to the next order in $c^2$.

\section{Calculating the spectral function}
\label{sec:spec}

In order to calculate the bulk spectral function, one needs to first
solve the equation of motion for the functions $f(z)$, $B(z)$ and
$\phi(z)$ and then use those results to solve Eq.~(\ref{eq:eom}) for
$H(z)$. This cannot be done analytically for an arbitrary potential
$V(\phi)$. However, the first step of this calculation can be done
analytically in the limit where the temperature is large compared with
the conformal breaking scale, $T \gg c^{1/\Delta_-}$. Another advantage of
this limit is that the results are universal in the sense that
the only property of the potential $V(\phi)$ which matters is the
curvature at the minimum, which determines the dimension of the
operator $\Delta_+$.

Equation (\ref{eq:eom}) would still need to be solved numerically,
however. We use a shooting procedure starting from the IR
boundary, i.e., the horizon $z=z_H$. The equation has a regular
singular point at $z=z_H$.
A Fr\"{o}benius expansion around this point can be
constructed iteratively. Among the two linearly independent solutions
the in-falling wave is chosen. The expansion is used to
calculate $H$ and $H'$ at a point near the IR boundary. These
values are then used as the initial conditions to integrate the
differential equation numerically all the way to the UV boundary $z=\varepsilon$.
The spectral function can then be determined from the numerical
results using Eq.~(\ref{eq:bulk}).

\begin{figure}[htb]
  \centering \subfigure[]{\label{fig:rhoa}
  \includegraphics[width=.45\textwidth]{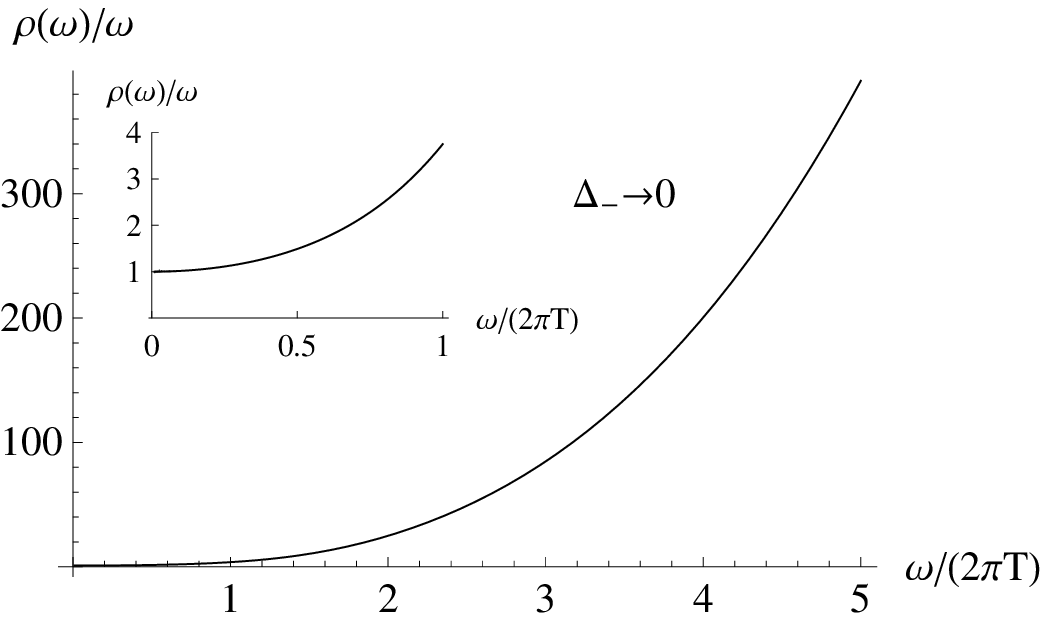}}
  \subfigure[]{\label{fig:rhob}
  \includegraphics[width=.45\textwidth]{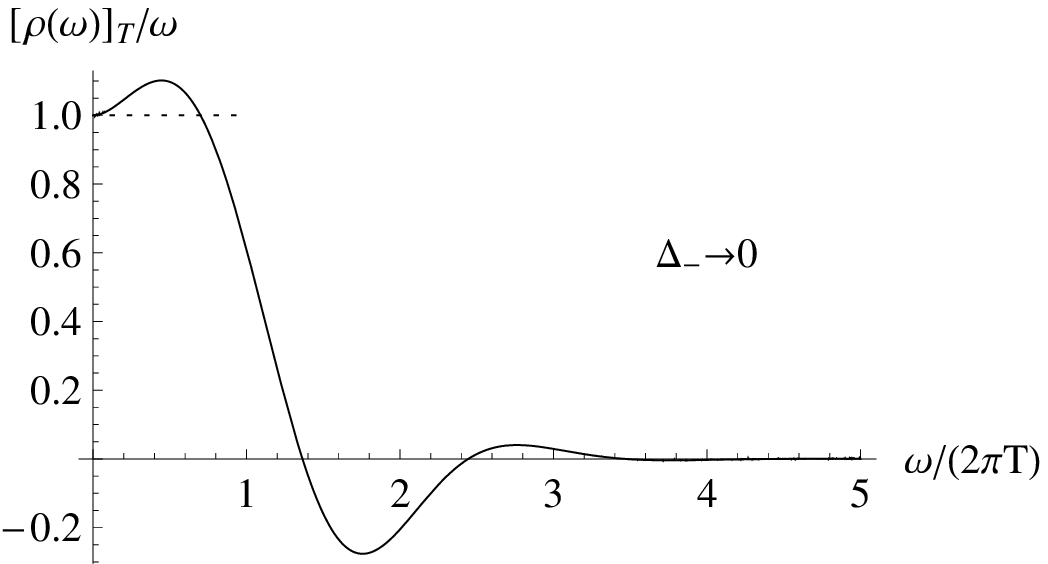}}
  \caption{The bulk spectral function divided by frequency in units of $\frac{1}{2\kappa^2} c^2 \Delta_-^2 (\pi T)^3$ for
    $\Delta_- \rightarrow 0$. a) The un-subtracted function.
   b) The function after the $T=0$ subtraction.}
  \label{fig:rho}
\end{figure}

\begin{figure}[htb]
  \centering
  \includegraphics[width=.45\textwidth]{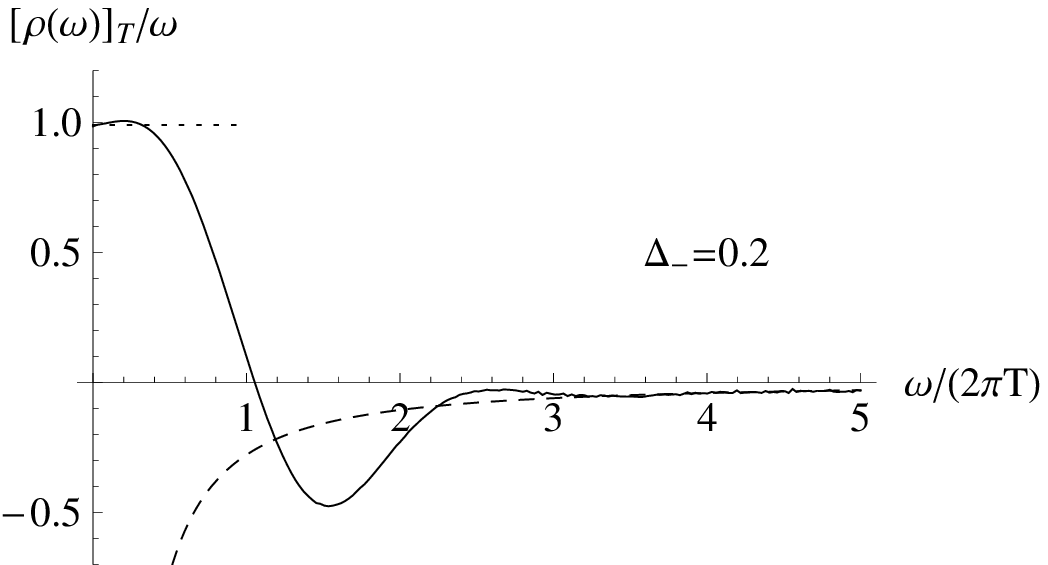}
  \includegraphics[width=.45\textwidth]{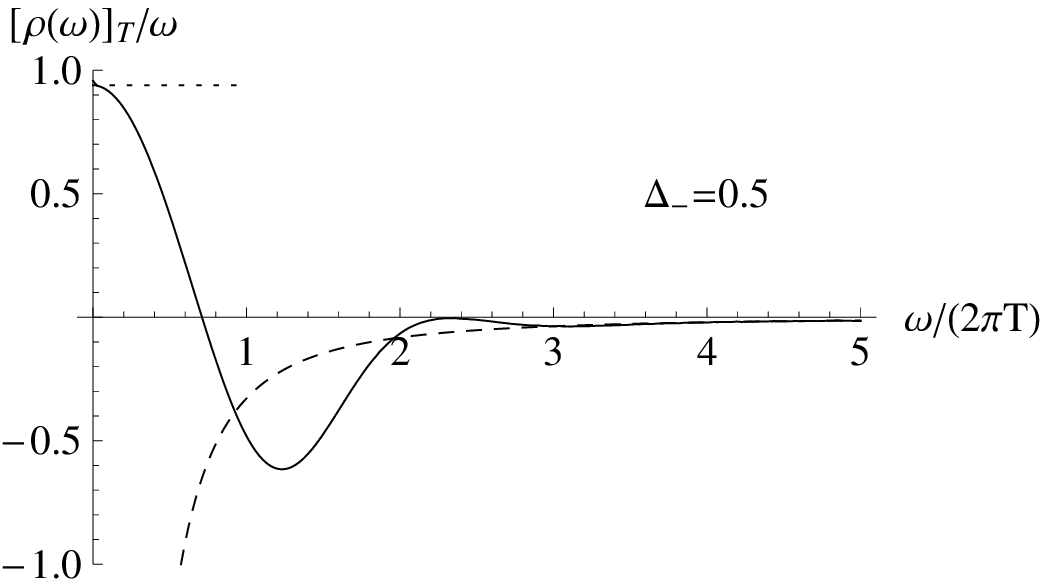}
  \includegraphics[width=.45\textwidth]{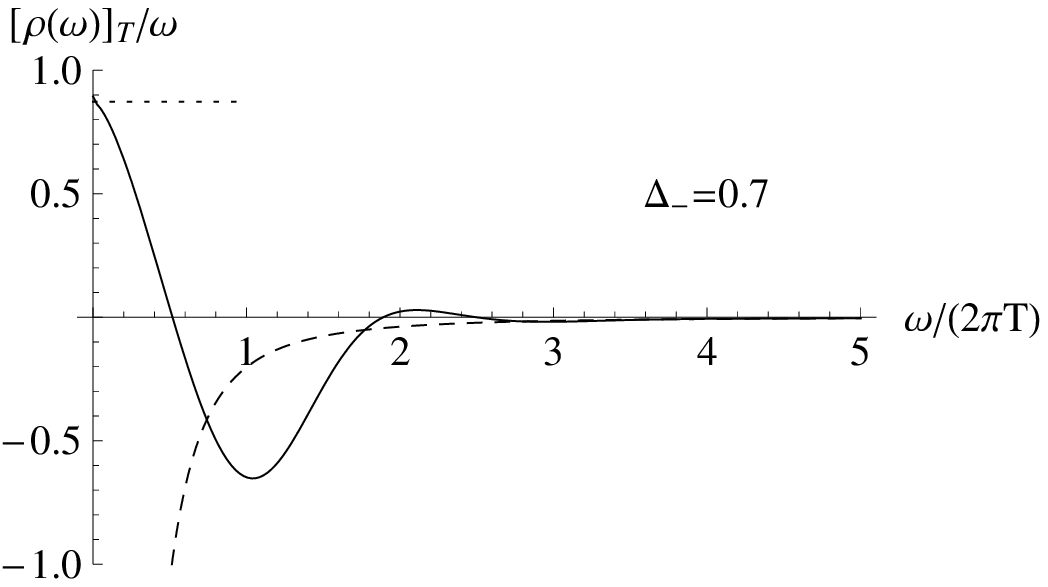}
  \includegraphics[width=.45\textwidth]{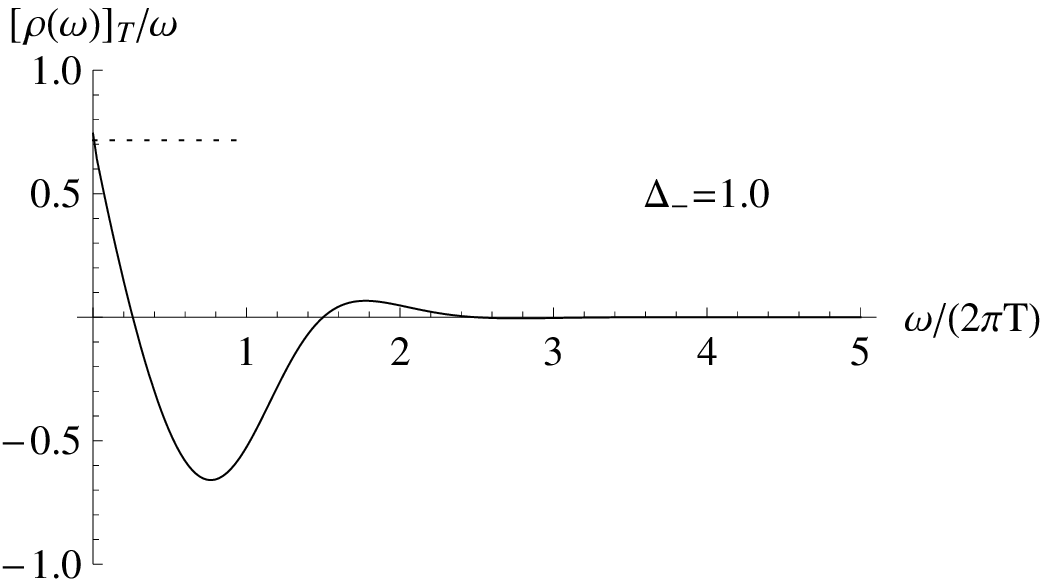}
  \caption{The bulk spectral function divided by frequency expressed
    in units of $\frac{1}{2\kappa^2} c^2 \Delta_-^2 (\pi
    T)^{3-2\Delta_-}$ for $\Delta_-$ equaling 0.2, 0.5, 0.7, and
    1.0. The dashed curve in each plot is from the analytic expression
    for the spectral function in the large $\omega$ limit in
    Eq.~(\ref{eq:rho1-rho0}).  The dotted line is drawn at the value
    of the bulk viscosity calculated from Eq.~(\ref{eq:bulk-visc}) in
    units of $\frac{1}{18 \kappa^2} c^2 \Delta_-^2 (\pi
    T)^{3-2\Delta_-}$. This value should be compared with the
    intercept of the plots at $\omega = 0$.  }
  \label{fig:drho}
\end{figure}

As a representative example, in Fig.~\ref{fig:rhoa}, we plot the
spectral function at order $c^2$ for $\Delta_-\to0$. As expected,
it is non-negative and it diverges for large $\omega$. The
subtracted spectral function $\deltaT{\rho(\omega)}$ to order
$\cO(c^2)$ is obtained by subtracting $\rho_0$ determined
analytically in Eq.~(\ref{eq:rho0}).
 Figures~\ref{fig:rhob} and
\ref{fig:drho} show $\deltaT{\rho(\omega)}/\omega$ plotted at
different values of $\Delta_-$, while the dashed line corresponds to
the analytic expressions for the large $\omega$ behavior given by
$\rho_1$ in Eq.~(\ref{eq:rho1-rho0}).

It is interesting to compare these plots with the lattice
calculations of the spectral function in Fig. 5 of
Ref.~\cite{Meyer:2010ii}. In both cases one finds the sign
oscillation of the spectral function in the region $\omega\sim
2\pi T$, which is absent in the weak coupling (Boltzmann equation)
result~\cite{Moore:2008ws,Hong:2010at}.

\section{Verifying the sum rule}
\label{sec:sum}

 In the high-temperature limit $T \gg c^{1/\Delta_-}$, the
left-hand side (thermodynamic side) of the sum
rule~Eq.~(\ref{eq:sum-rule-Dplus})
 can be analytically calculated in the holographic model. The trace
anomaly is given by Eq.~(\ref{eq:tanom}), while the derivative
with respect to entropy can be expressed via derivative with
respect to enthalpy $w$. Using the expression for $d$ in the high $T$
limit, Eq.~(\ref{eq:dc}), the derivative can be performed, and the
left-hand side can be expressed as,
\begin{equation} \label{eq:lhs}
\left( 3 s \frac{\partial}{\partial s} -\Delta_+\right) \left( \epsilon -
3 p \right) = \frac{c\, d}{2\kappa^2}\, \Delta_-^2
\left(\Delta_+-\Delta_-\right).
\end{equation}
The right-hand side can be numerically calculated from
$\deltaT{\rho}$. For efficiency, we make use of the analytical result for the
asymptotics of $\deltaT{\rho}$ Eq.~(\ref{eq:rho1-rho0}). We
calculate numerically only the contribution to the integral up to some
relatively large frequency $\omega_{\rm max}$. The contribution from
$\omega_{\rm max}$ to $\infty$ is calculated analytically using asymptotics
$\deltaT{\rho}\approx\rho_1$ from Eq.~(\ref{eq:rho1-rho0}). We chose
$\omega_{\rm max} = 10\pi T$. Table~\ref{tab:sum} shows that the sum rule
holds to at least within a fraction of a percent.
\begin{table}[htb]
\begin{tabular}{c|c|c|c}
 $\Delta_-$& LHS&RHS&$\%$
Error\\
\hline 1.0&-0.4569&-0.4588&0.4 \\
0.7&-0.4862&-0.4889&0.6\\
0.5&-0.4951&-0.4984&0.6\\
0.2&-0.4997&-0.5025&0.5\\ \hline
\end{tabular}
\caption{Verification of the sum rule in Eq.~(\ref{eq:sum-rule-Dplus}) in the
  holographic model for a sample of values of $\Delta_-$ in the
  high-temperature limit $T\gg c^{1/\Delta_-}$. The columns LHS
  and the RHS are the values of the left- and the right- hand side of
  Eq.~(\ref{eq:sum-rule-Dplus}) expressed in units of
  $\frac{1}{2\kappa^2} c^2 \Delta_-^3 (\pi T)^{3-2\Delta_-}$, which
  removes the dependence on $\kappa$, $c$, $T$ and the leading
  ($\Delta_-^3$) dependence on $\Delta_-$.
  The LHS is calculated analytically using Eq.~(\ref{eq:lhs}), while the RHS is a numerical
  calculation of the integral as described in text. The last column
  shows the
  percent discrepancy between the two sides.}\label{tab:sum}
\end{table}

\section{The marginal limit and the tail of the spectral function}
\label{sec:tail}

As $\Delta_-\to 0$ the left-hand side of the sum rule given by
Eq.~(\ref{eq:lhs}) scales as $\Delta_-^3$ ($d\sim\Delta_-$ according to
Eq.~(\ref{eq:dc})). This is also apparent in
Table~\ref{tab:sum}.

However, the spectral function $\deltaT{\rho}$ on the right-hand
side scales as $\Delta_-^2$.  This comes from factor $B'(z)$ in
Eq.~(\ref{eq:bulk}) which scales as $\Delta_-^2$ according to
Eq.~(\ref{eq:b-small-z}). Therefore, in order to determine the
$\cO(\Delta_-^2)$ part of $\rho(\omega)$ we can simply set $\Delta_-=0$ in
Eq.~(\ref{eq:eom}) which then becomes~\footnote{Note that according to Eq.~(\ref{eq:b-high-t}), $B'(z)
\sim c^2$, which means, to $\cO(c^2)$ we are working, we can drop
terms like $B'(z)$, but not terms like
$B''(z)/B'(z) \to -1/z$.}
\begin{equation} \label{eq:eom-small-d}
H''(z)=H'(z)\left(\frac{3}{z}-\frac{f'(z)}{f(z)}\right)-\frac{\omega^2
}{f(z)^2} H(z)\,,
\end{equation}
where $f$ is given by Eq.~(\ref{eq:fz}) in the high-temperature limit.
 This is the well known equation of motion of a massless
scalar field on a pure AdS background. However, the analytic
solution to this equation with required boundary conditions has
not yet been found (see
\cite{Starinets:2002br,Nunez:2003eq,Myers:2007we,Berti:2009kk} for
discussion). Nevertheless, the solution can be obtained
numerically as discussed in Sec.~\ref{sec:sum}. The resulting
spectral function is displayed in Fig.~\ref{fig:rho}. The
$\cO(\Delta_-^2)$ part of $\deltaT{\rho(\omega)}$ shown in
Fig.~\ref{fig:rhob} oscillates around 0, but the contribution from
it to the spectral integral in the sum rule is non-vanishing. We
calculate the integral numerically and find
\begin{equation} \label{eq:smalld}
\frac{2}{\pi} \int_0^{\omega_{\rm max}}
 \frac{\deltaT{\rho(\omega)}}{\omega} \,d \omega
= 2 c^2 \Delta_-^2 \left(\frac{w}{4}\right) \left(.60000(1)\right)
+\cO(\Delta_-^3)\,,
\end{equation}
where $w$ is the enthalpy. The integral of the $\Delta_-^2$ part of $\deltaT{\rho(\omega)}$ is
converging very fast (apparently exponentially) at $\omega=\infty$ and
we picked relatively large $\omega_{\rm max}=10\pi T$ for this numerical
calculation.  The number in
parenthesis is $3/5$ to at least four significant digits.
How then is the sum rule satisfied if the left-hand side is only $\cO(\Delta_-^3)$?

We shall find the missing contribution in the tail of the function
$\deltaT{\rho(\omega)}$. Indeed,  taking $\omega_{\rm max}$ to $\infty$ requires
extra care. Because $\cO(\Delta_-^2)$ part of $\deltaT{\rho(\omega)}$
decreases very fast as $\omega\to\infty$, for sufficiently large $\omega$
the dominant part in $\deltaT{\rho(\omega)}$ is of order $\Delta_-^3$.
As we see analytically in
Eqs.~(\ref{eq:rho1-rho0}),~(\ref{eq:rho0}),
$\deltaT{\rho}\to\rho_1\sim\Delta_-^3\omega^{-2\Delta_-}$. At any finite
$\Delta_-$ the negative power-law  tail effectively cuts off  the integral at large
$\omega_{\rm tail}\sim\exp(1/(2\Delta_-))$. However, the contribution
of this long tail
to the integral grows with $\omega_{\rm tail}$ as
$\int_{\omega_{\rm max}}^{\omega_{\rm tail}}d\omega\,\deltaT{\rho}/\omega\sim
\Delta_-^3\log(\omega_{\rm tail}/\omega_{\rm
  max})\sim\Delta_-^2$, i.e., the length of the tail compensates for
the extra power of $\Delta_-$.

Let us calculate this $\cO(\Delta_-^2)$ contribution from the tail. Using
Eqs.~(\ref{eq:rho1-rho0}),~(\ref{eq:rho0}) we can write for large
$\omega$ and small~$\Delta_-$:
\begin{equation} \label{eq:rholargew}
\deltaT{\rho(\omega)}= \rho_1(\omega)+\ldots
= -\frac{6}{5} \pi c^2 \Delta_-^3 \omega^{-2\Delta_-}
\left(\frac{w}{4}\right)+\ldots.
\end{equation}
Thus
\begin{equation}
\frac{2}{\pi} \int_{\omega_{\rm max}}^\infty \frac{{ d\omega}}{\omega}\,\deltaT{\rho} = - 2 c^2
\Delta_-^2 \left( \frac{w}{4}\right) \left(\frac{3}{5}\right)+\cO(\Delta_-^3).
\end{equation}
We see that the contribution of the high-frequency tail exactly
cancels the contribution from the region of $\omega\sim T$
in Eq.~(\ref{eq:smalld}). Therefore the RHS
of the sum rule is proportional to $\Delta_-^3$ just as the
LHS.

Remarkably, this cancellation mechanism is very similar to the one
which was found in QCD by Caron-Huot in Ref.\cite{CaronHuot:2009ns}.
In the case of QCD the left-hand (thermodynamic) side of the sum rule
is of order $\alpha_s^3$, while the spectral function on the
right-hand side is $\cO(\alpha_s^2)$, with logarithmically long
high-frequency tail
of order $\cO(\alpha_s^3)$.

We would like also to comment that the $\Delta_-^2$ contribution to the
bulk spectral function $\deltaT{\rho}$ in the high-temperature regime is
the same, up to a constant, as the spectral function in the $N=4$ SUSY
YM theory. In fact, the spectral integral in Eq.~(\ref{eq:smalld}) has been
also performed numerically in Ref.~\cite{Romatschke:2009ng}, using a
different method and with
the same result. Unlike the case we consider, however, for
the shear channel sum rule considered in
Ref.~\cite{Romatschke:2009ng} that integral saturated the sum rule.

Also worth noting is that the shear and the bulk channel spectral
functions are proportional to each other in the Chamblin-Real dilaton
model examined in \cite{Springer:2010mw} and
\cite{Springer:2010mf}. In such theories the mechanism of the
saturation of the bulk sum rule is the same as that of the shear sum rule
and no high/low frequency cancellations are needed in either case.

\section{Viscosity}
\label{sec:viscosity}

As another cross-check of our results, we can compare them to the
analytic calculation of the bulk viscosity in the high temperature
limit of Ref.~\cite{Yarom:2009mw}, which finds
\begin{equation} \label{eq:bulk-visc}
\zeta = \frac{1}{9}
\frac{c^2 \Delta_-^2 }{2\kappa^2}\,(\pi T)^{3-2\Delta_-}\,
2^{\Delta_-}\pi\,
\left(
\frac{\Gamma(1-\Delta_-/4)}{\Gamma(1/2-\Delta_-/4)}\right)^2.
\end{equation}
This formula was derived by matching gradient expansion of the
stress-energy tensor to the gradient expansion of the background
metric.

Ideally, one should also be able to derive Eq.~(\ref{eq:bulk-visc}) by
solving Eq.~(\ref{eq:eom}) at small $\omega$ and using Kubo formula
\begin{equation}
  \label{eq:zeta-rho}
  \zeta = \frac19\rho'(0).
\end{equation}
Indeed, this can be done analytically for $\Delta_- \rightarrow
0$. In this case, Eq.~(\ref{eq:eom}) reduces to
Eq.~(\ref{eq:eom-small-d}). For small $\omega$, this equation can
be easily integrated. Normalizing as $H(0)=1$ and using in-falling
boundary condition at the horizon we find~\footnote{Although this
solution is not valid all the way to the horizon due to the
singularity, for sufficiently small $\omega$ it is valid close
enough to the horizon to allow matching it to the in-falling wave
solution.}
\begin{equation}
H(z) = 1-\frac{i\, \omega}{4}\left(\frac{\bar w}{4}\right)^{-1/4}
\log(1-\bar w z^4/4)+\ldots\,.
\end{equation}
This gives for the spectral function at small $\omega$,
\begin{equation}
\rho = \frac{c^2 \Delta_-^2 }{2\kappa^2}\,(\pi T)^3\,\omega +\ldots,
\end{equation}
and for the bulk viscosity
\begin{equation}
\zeta = \frac{1}{9} \frac{c^2 \Delta_-^2 }{2\kappa^2}\,(\pi T)^3.
\end{equation}
This agrees with the $\Delta_- \to
0$ limit of  Eq.~(\ref{eq:bulk-visc}).

Unfortunately, a similar approach at finite $\Delta_-$ appears
intractable, since an analytic solution to Eq.~(\ref{eq:eom}),
even at small $\omega$, for arbitrary $\Delta_-$ is not known.
However, the bulk viscosity can be calculated  by applying
the Kubo formula Eq.~(\ref{eq:zeta-rho}) to our numerical results for the
spectral function. We have verified the agreement of
Eq.~(\ref{eq:bulk-visc}) with these numerical calculations and
illustrated it in Fig.~\ref{fig:drho}.

\section{Conclusion}
\label{sec:conc}

We studied the spectral function corresponding to time-dependent
bulk deformation (uniform expansion) in a class of field theories
where conformality is broken ``softly'' in the sense that at high
temperature the equation of state approaches conformal limit
$\epsilon=3p$, similar to QCD. The breaking is due to a scalar
operator of conformal dimension $\Delta_+<4$, which is an analogue
of the operator of gluon condensate in QCD. We find that in such
theories the bulk spectral function satisfies the sum rule given
by Eq.~(\ref{eq:sum-rule-Dplus}).

The sum rule in Eq.~(\ref{eq:sum-rule-Dplus}) is similar to the
sum rule in an asymptotically free theory such as QCD derived by
Romatschke and Son in Ref.~\cite{Romatschke:2009ng}. In the
marginal limit $\Delta_+\to4$, the two sum rules are identical. We
used this similarity to address an interesting puzzle noted in
Ref.~\cite{Moore:2008ws}. In order to satisfy the sum rule, a
delicate cancellation must occur between the regions of
$\omega\sim T$ and of $\omega\gg T$ in the spectral integral.
Ref.\cite{CaronHuot:2009ns} showed that this cancellation can be
indeed seen in the high-temperature limit of QCD, using weak
coupling calculation. Our study of the holographic model shows
that such a cancellation is very generic to the whole class of
strongly coupled theories with softly broken conformal symmetry.

To make the connection of our model with QCD more tangible, we can
observe that in QCD the action contains operator
$G_{\mu\nu}G^{\mu\nu}/\alpha_s$ and thus identify the operator $\oO$
with $G_{\mu\nu}G^{\mu\nu}/\alpha_s$, up to a numerical
coefficient. This coefficient can be determined by matching
correlation functions (e.g., the bulk spectral function) of the
holographic model to QCD, but we shall not need it here. The anomalous
dimension of this operator in QCD is given by
$\beta(\alpha_s)/\alpha_s=b_0\alpha_s/(2\pi)$. In QCD this anomalous
dimension is a function of scale, vanishing logarithmically, i.e.,
very slowly, with increasing energy-momentum scale.
In the class of QCD-like theories we consider, the corresponding
quantity is $\Delta_+-4=-\Delta_-$, which is a constant.\footnote{
Holographic models which reproduce the effect of the logarithmic
running of $\alpha_s$, have been discussed in, e.g., Refs.\cite{Csaki:2006ji,Gursoy:2007cb,Gursoy:2007er}.
}
 This
scaling dimension taken at $\omega=\infty$, determines the value of
$\deltaT{G_R(i\infty)}$ in the sum rule~(\ref{eq:sum-rule}), according
to Eq.~(\ref{eq:GR-infty}), and it is responsible for the difference
of the general sum rule~(\ref{eq:sum-rule-Dplus}) from QCD.

The holographic theories we consider can serve only as a
qualitative or semi-quantitative guide to the QCD phenomena, since
QCD becomes a weakly coupled theory at sufficiently high
energy-momentum scale due to asymptotic freedom.  But this guide
might be useful by offering a view complementary to the weak
coupling extrapolation out of the domain of asymptotically high
energies. The experiments at RHIC provide a powerful argument that
the domain of interest in heavy-ion collisions is a strongly
coupled domain. It will be very interesting to see to what extent
this remains true at LHC energies.

Also notable in this regard are lattice calculations of the spectral
function. The striking feature of the lattice results is the sign
oscillation of the spectral function in the $\omega\sim 2\pi T$
region. This oscillation is absent in weakly coupled
calculations~\cite{CaronHuot:2009ns,Hong:2010at},
but appears to be a generic feature in the
spectral functions obtained in holographic models, as our results in
Figs.~\ref{fig:rhob} and \ref{fig:drho} illustrate. This qualitative
difference appears to be another manifestation of the now familiar fact that
real-time response (in particular, hydrodynamics) is more sensitive to
the coupling strength than, e.g., equation of state.

\acknowledgments

 This work is supported by the DOE grant
No.\ DE-FG0201ER41195.

\begin{appendix}

\section{Background of the holographic model}
\label{sec:appA}

\subsection{Einstein's equations and one-point functions}

This appendix summarizes, for completeness, the relevant results of
Ref.~\cite{Hohler:2009tv}.
The background geometry corresponding to homogeneous
boundary conditions on $g_{\mu\nu}$ and $\phi$ can be found by solving
Einstein's equations for  the metric in
Eq.~(\ref{eq:metric}) which take the form:
\begin{eqnarray}
&\dot{B} = -\frac{1}{6}\, \dot{\phi}^2, \label{eq:ein1}\\
&\ddot{f} = \left({4} + \dot{B}\right) \dot{f}, \label{eq:ein2}\\
&-6 \dot{f} + f \left({24} - \dot{\phi}^2\right) + 2 {e^{2 B}}\, V(\phi) = 0, \label{eq:ein3}\\
&\ddot{\phi}f + \dot{\phi}\left(\dot{f} - f(4 + \dot{B})\right) -
e^{2 B}\, dV(\phi)/d\phi = 0, \label{eq:ein4}
\end{eqnarray}
where a dot denotes a $\log z$ derivative, {\it e.g.},
$\dot\phi=z\,d\phi/dz$. The holographic
correspondence provides the boundary conditions at the UV boundary
$z=\varepsilon$ given by Eqs.~(\ref{eq:b-c}).
Minkowski metric at the boundary requires
\begin{equation}
  \label{eq:f-bc}
  f(\varepsilon)=1\,.
\end{equation}
Equation (\ref{eq:ein2}) can be integrated once to give
\begin{equation} \label{eq:ein2b}
\dot{f} = - \bar w z^4 e^{B}.
\end{equation}
The integration constant $\bar w$ must be positive if the metric is to
possess a horizon $f(z_H)=0$ at some value of $z_H$. Since $\bar w$
determines the position of the horizon, it is related to temperature,
and we find that it is proportional to the enthalpy,
Eq.~(\ref{eq:w-bar}).

A boundary condition on $B$ is not needed because the
value of $B$ is determined by Eq.~(\ref{eq:ein3}), which is
algebraic in $B$. The role of the second boundary condition for
Eq.~(\ref{eq:ein4}) is played by the requirement that $\phi$ is
finite at the horizon, $z=z_H$, which is a regular singular point
of the second order differential equation~(\ref{eq:ein4}).

Near the $z=\varepsilon\to0$ boundary, $\phi\to0$, $B\to0$, $\dot
B\to 0$ and $\dot f\to 0$.  Equation~(\ref{eq:ein4}) for $\phi$
can be linearized and the asymptotic behavior of $\phi$ near the
boundary can be determined easily:
\begin{equation} \label{eq:asy}
\phi(z) \to (c - d\,
\varepsilon^{\Delta_+-\Delta_-})\,z^{\Delta_-}(1 + \ldots)
+ d\, z^{\Delta_+}
(1 + \ldots),
\end{equation}
where the curvature of the potential $V''(0)\equiv m_5^2$ determines
the indices $\Delta_\pm=2\pm\sqrt{4+m^2}$. The coefficient of the
first term is related to $c$ by Eq.~(\ref{eq:b-c}). The
coefficient $d$ of the second linearly independent solution should
be determined by the finiteness condition at the horizon and is a
function of $\bar w$ (i.e., temperature) and $c$.

By calculating the derivative of the 5D action with respect to $c$
and matching it, by holographic correspondence, to the expectation
value $\langle{\oO}\rangle$, one finds (see also \cite{Klebanov:1999tb})
\begin{equation} \label{eq:op}
\langle \oO \rangle = -\frac{\partial S_5}{\partial c} =
-\frac{e^{-B(z)}\phi'(z)}{2\kappa^2z^{3-\Delta_-}}\Big|_{z=\epsilon}
=
-\frac{d}{2\kappa^2} \, (\Delta_+-\Delta_-)
+ \ldots,
\end{equation}
where ``\ldots'' denote UV
divergent but temperature independent terms.
One can thus see that the integration constant $d$ is related to
the UV finite (and temperature dependent) part of $\langle\oO\rangle$.

Furthermore, from the expression for $\phi(z)$ near the boundary Eq.~(\ref{eq:asy})
and Eq.~(\ref{eq:ein1}), the function $\dot{B}(z)$
can be calculated for small $z$:
\begin{equation}\label{eq:b-small-z}
\dot{B}(z)= zB'(z) = -\frac{1}{6} \Delta_-^2 \left(c-d
\varepsilon^{\Delta_+-\Delta_-}\right)^2 z^{2\Delta_-} -
\frac{1}{3} \Delta_-\Delta_+
\left(c-d\varepsilon^{\Delta_+-\Delta_-}\right) d
z^{\Delta_++\Delta_-} \ldots.
\end{equation}

By considering homogeneous variations of the boundary condition on
the metric, the one-point functions of the stress-energy
tensor can be calculated using Eq.~(\ref{eq:Z-g-Tmunu}) as in
Ref.~\cite{Hohler:2009tv}.
\begin{align}
\langle T^{00}\rangle = \left.-\frac{6
e^{-B(z)}}{2 \kappa^2 z^4}\right|_{z=\varepsilon},\qquad
\langle T^{11} \rangle = \frac{\bar w}{2\kappa^2}-\langle T^{00}\rangle.
\end{align}
The thermal energy and pressure,
\begin{align} \label{eq:e-p}
\epsilon = \langle T^{00}\rangle - \langle T^{00}\rangle_{T=0}
\equiv\deltaT{\langle T^{00}\rangle},\qquad
p = \langle T^{11}\rangle - \langle T^{11}\rangle_{T=0}
\equiv\deltaT{\langle T^{11}\rangle},
\end{align}
are finite at $\varepsilon=0$ and equal to zero at $T=0$. The
enthalpy, $w=\epsilon+p$ is related to the constant $\bar w$:
\begin{equation}
w=\frac{\bar w}{2\kappa^2}\label{eq:w-bar}.
\end{equation}
 After solving
Eq.~(\ref{eq:ein3}) for $B$ at $z=\varepsilon$ with $\phi$ given
by Eq.~(\ref{eq:asy}), we find that the energy and pressure can be
expressed as
\begin{align}
\epsilon &=
\frac{w}{4}-\frac{c\,\deltaT{d}}{8\kappa^2}\,
\Delta_-(\Delta_+-\Delta_-),\qquad
p = w - \epsilon.
\end{align}
Therefore the temperature dependence of the expectation value of the
trace anomaly is given by,
\begin{equation}\label{eq:tanom}
\deltaT{\langle \theta \rangle} =  3p-\epsilon = \frac{c\,\deltaT{d}\,\Delta_-}{2
\kappa^2}\,(\Delta_+-\Delta_-) = -\Delta_-\, c\, \deltaT{\langle \oO
\rangle}.
\end{equation}
in accordance with the anomaly equation~(\ref{eq:theta-O}).
\subsection{High temperature limit}

The equations of motion can be simplified and analytically
solved if one considers the high temperature limit, or the limit
where $T \gg c^{1/\Delta_-}$. Since the enthalpy is related to the
temperature,
\begin{equation}
  \label{eq:w-T}
  \bar w =4\left(\pi T\right)^4
\left(1 + \cO\left(c^2T^{-2\Delta_-}\right)\right),
\end{equation}
 this limit can also be expressed as $\bar w \gg
c^{4/\Delta_-}$.
One can begin by observing that at large $\bar w$ the
function $f$ varies very rapidly according to
Eq.~(\ref{eq:ein2b}). This means one can neglect variation of the
function $B$ between the boundary $z=\varepsilon$ and the horizon
$z=z_H$, since $z_H$ becomes small (as $ \bar w^{-1/4}$). Since on the
boundary $B=0$ (up to terms of order $\varepsilon^{2\Delta_-}$,
negligible here, according to Eq.~(\ref{eq:ein3})), we find from
Eq.~(\ref{eq:ein2b})
\begin{equation}
  \label{eq:fz}
   f(z) = 1 - \bar w\, z^4/4.
\end{equation}
Another consequence is that $\phi$, which is small at
$z=\varepsilon$, remains small up to $z_H$ ($\phi\sim
cz_H^{\Delta_-}\sim c/T^{\Delta_-}\ll1$), and the linearized
approximation to Eq.~(\ref{eq:ein4}) is valid not only near the
boundary, but all the way to the horizon. With $B=0$ and $f$ from
Eq.~(\ref{eq:fz}) we obtain
\begin{equation} \label{eq:dphi}
\left(1 - \frac{1}{4}\, \bar w z^4\right) {\phi}'' - \left (
\frac{3}{z} + \frac{\bar w z^3}{4} \right) {\phi}' -
\frac{m^2}{z^2}\,\phi = 0.
\end{equation}
Equation (\ref{eq:dphi}) can be solved analytically
\begin{equation} \label{eq:phi}
\begin{split}
\phi(z) & = c\,z^{\Dm}\;
{}_2F_1\left(\hfrac{\Dm}{4},\,\hfrac{\Dm}{4},\,\hfrac{\Dm}{2},\,\bar w z^4\right/4)\\
& + d \, z^{\Dp}
\;{}_2F_1\left(\hfrac{\Dp}{4},\,\hfrac{\Dp}{4},\,\hfrac{\Dp}{2},\,
\bar w z^4/4\right),
\end{split}
\end{equation}
where the coefficients follow the notations of Eq.~(\ref{eq:asy})
(up to terms ${\cal
  O}(\varepsilon^{\Delta_+-\Delta_-})$, here negligible). Both linearly independent solutions are
logarithmically divergent at the horizon $z=z_H$, where
$\bar wz_H^4/4=1$. The condition $|\phi(z_H)|<\infty$ requires us to
select the linear combination in which these divergences cancel.
This fixes $d$ in terms of $c$:
\begin{equation} \label{eq:dc}
\begin{split}
d &= - c\; \bar w^{{(\Dp-\Dm)}/{4}}\, D(\Dm)\,,\\
\end{split}
\end{equation}
where the function $D(\Dm)=1/D(\Dp)$ is given by
\begin{equation} \label{eq:DD}
\begin{split}
D(\Dm)
&=\frac{\pi \,2^{\Dm}}{2-\Dm} \cot\left(\pi \Dm/4\right)
\frac{\Gamma(\Dm/2)^2}{\Gamma(\Dm/4)^4}
\,.
\end{split}
\end{equation}
This solution for $\phi(z)$ can be used in Eq.~(\ref{eq:ein1}) to
calculate $B^\prime(z)$,
\begin{equation}
\begin{split} \label{eq:b-high-t}
B'(z) &= -\frac{1}{6} \left(  c \,\Delta_- z^{\Delta_-+1}
{}_2F_1\left(\Delta_-/4,\,1+\Delta_-/4,\,\Delta_-/2,\, \bar w
z^4/4\right)\right. \\ & \qquad + \left.d \,\Delta_+
z^{\Delta_++1} {}_2F_1
\left(\Delta_+/4,\,1+\Delta_+/4,\,\Delta_+/2,\, \bar w
z^4/4\right)\right)^2
\end{split}
\end{equation}

From this expression, the function $B(z)$ can be calculated to order
$c^2$, which is the leading order in the high-temperature limit.
By iteratively solving the equations of motion, the
higher order corrections to $B$ as well as to $f$ and $\phi$ can be
determined if necessary.

\end{appendix}


\begin{thebibliography}{99}



\bibitem{Adcox:2004mh}
  K.~Adcox {\it et al.} [ PHENIX Collaboration ],
  Nucl.\ Phys.\  {\bf A757}, 184-283 (2005).
  [nucl-ex/0410003].

\bibitem{Back:2004je}
  B.~B.~Back, M.~D.~Baker, M.~Ballintijn {\it et al.},
  Nucl.\ Phys.\  {\bf A757}, 28-101 (2005).
  [nucl-ex/0410022].

\bibitem{Arsene:2004fa}
  I.~Arsene {\it et al.} [ BRAHMS Collaboration ],
  Nucl.\ Phys.\  {\bf A757}, 1-27 (2005).
  [nucl-ex/0410020].

\bibitem{Adams:2005dq}
  J.~Adams {\it et al.} [ STAR Collaboration ],
  Nucl.\ Phys.\  {\bf A757}, 102-183 (2005).
  [nucl-ex/0501009].



\bibitem{Nakamura:2004sy}
  A.~Nakamura, S.~Sakai,
  Phys.\ Rev.\ Lett.\  {\bf 94}, 072305 (2005).
  [hep-lat/0406009].

\bibitem{Aarts:2007wj}
  G.~Aarts, C.~Allton, J.~Foley {\it et al.},
  Phys.\ Rev.\ Lett.\  {\bf 99}, 022002 (2007).
  [hep-lat/0703008 [HEP-LAT]].

\bibitem{Meyer:2007ic}
  H.~B.~Meyer,
  Phys.\ Rev.\  {\bf D76}, 101701 (2007).
  [arXiv:0704.1801 [hep-lat]].


\bibitem{Meyer:2007dy}
  H.~B.~Meyer,
  Phys.\ Rev.\ Lett.\  {\bf 100}, 162001 (2008).
  [arXiv:0710.3717 [hep-lat]].

\bibitem{Huebner:2008as}
  K.~Huebner, F.~Karsch, C.~Pica,
  Phys.\ Rev.\  {\bf D78}, 094501 (2008).
  [arXiv:0808.1127 [hep-lat]].




\bibitem{Meyer:2008sn}
  H.~B.~Meyer,
  PoS {\bf LATTICE2008}, 017 (2008)
  [arXiv:0809.5202 [hep-lat]].

\bibitem{Iqbal:2009xz}
  N.~Iqbal and H.~B.~Meyer,
  JHEP {\bf 0911}, 029 (2009)
  [arXiv:0909.0582 [hep-lat]].

\bibitem{Meyer:2010ii}
  H.~B.~Meyer,
  JHEP {\bf 1004}, 099 (2010)
  [arXiv:1002.3343 [hep-lat]].



\bibitem{Kharzeev:2007wb}
  D.~Kharzeev and K.~Tuchin,
  JHEP {\bf 0809}, 093 (2008)
  [arXiv:0705.4280 [hep-ph]].

\bibitem{Karsch:2007jc}
  F.~Karsch, D.~Kharzeev and K.~Tuchin,
  Phys.\ Lett.\  B {\bf 663}, 217 (2008)
  [arXiv:0711.0914 [hep-ph]].


\bibitem{Romatschke:2009ng}
  P.~Romatschke, D.~T.~Son,
  Phys.\ Rev.\  {\bf D80}, 065021 (2009).
  [arXiv:0903.3946 [hep-ph]].



\bibitem{Maldacena:1997re}
  J.~M.~Maldacena,
  Adv.\ Theor.\ Math.\ Phys.\  {\bf 2}, 231 (1998)
  [Int.\ J.\ Theor.\ Phys.\  {\bf 38}, 1113 (1999)];
%

\bibitem{Gubser:1998bc}
  S.~S.~Gubser, I.~R.~Klebanov and A.~M.~Polyakov,
  Phys.\ Lett.\ B {\bf 428}, 105 (1998);

\bibitem{Witten:1998qj}
E.~Witten,
  Adv.\ Theor.\ Math.\ Phys.\  {\bf 2}, 253 (1998).



\bibitem{Aharony:1999ti}
  O.~Aharony, S.~S.~Gubser, J.~M.~Maldacena, H.~Ooguri and Y.~Oz,
  Phys.\ Rept.\  {\bf 323}, 183 (2000)
  [arXiv:hep-th/9905111].

\bibitem{Son:2007vk}
  D.~T.~Son, A.~O.~Starinets,
  Ann.\ Rev.\ Nucl.\ Part.\ Sci.\  {\bf 57}, 95-118 (2007).
  [arXiv:0704.0240 [hep-th]].


\bibitem{Erdmenger:2007cm}
  J.~Erdmenger, N.~Evans, I.~Kirsch and E.~Threlfall,
  Eur.\ Phys.\ J.\  A {\bf 35}, 81 (2008)
  [arXiv:0711.4467 [hep-th]].


\bibitem{Myers:2008fv}
  R.~C.~Myers and S.~E.~Vazquez,
  Class.\ Quant.\ Grav.\  {\bf 25}, 114008 (2008)
  [arXiv:0804.2423 [hep-th]].

\bibitem{Gubser:2009md}
  S.~S.~Gubser and A.~Karch,
  Ann.\ Rev.\ Nucl.\ Part.\ Sci.\  {\bf 59}, 145 (2009)
  [arXiv:0901.0935 [hep-th]].


\bibitem{Ammon:2010zz}
  M.~Ammon,
  Fortsch.\ Phys.\  {\bf 58}, 1123 (2010).

\bibitem{CasalderreySolana:2011us}
  J.~Casalderrey-Solana, H.~Liu, D.~Mateos, K.~Rajagopal, U.~A.~Wiedemann,
    [arXiv:1101.0618 [hep-th]].





\bibitem{Gubser:2008ny}
  S.~S.~Gubser, A.~Nellore,
  Phys.\ Rev.\  {\bf D78}, 086007 (2008).
  [arXiv:0804.0434 [hep-th]].

\bibitem{Gubser:2008yx}
  S.~S.~Gubser, A.~Nellore, S.~S.~Pufu {\it et al.},
  Phys.\ Rev.\ Lett.\  {\bf 101}, 131601 (2008).
  [arXiv:0804.1950 [hep-th]].

\bibitem{Gubser:2008sz}
  S.~S.~Gubser, S.~S.~Pufu, F.~D.~Rocha,
  JHEP {\bf 0808}, 085 (2008).
  [arXiv:0806.0407 [hep-th]].



\bibitem{Hohler:2009tv}
  P.~M.~Hohler, M.~A.~Stephanov,
  Phys.\ Rev.\  {\bf D80}, 066002 (2009).
  [arXiv:0905.0900 [hep-th]].

\bibitem{Cherman:2009tw}
  A.~Cherman, T.~D.~Cohen and A.~Nellore,
  Phys.\ Rev.\  D {\bf 80}, 066003 (2009)
  [arXiv:0905.0903 [hep-th]].



\bibitem{Springer:2010mw}
  T.~Springer, C.~Gale, S.~Jeon,
  [arXiv:1010.2760 [hep-th]].


\bibitem{Chamblin:1999ya}
  H.~A.~Chamblin, H.~S.~Reall,
  Nucl.\ Phys.\  {\bf B562}, 133-157 (1999).
  [hep-th/9903225].





\bibitem{Moore:2008ws}
  G.~D.~Moore, O.~Saremi,
  JHEP {\bf 0809}, 015 (2008).
  [arXiv:0805.4201 [hep-ph]].

\bibitem{CaronHuot:2009ns}
  S.~Caron-Huot,
  Phys.\ Rev.\  {\bf D79}, 125009 (2009).
  [arXiv:0903.3958 [hep-ph]].


\bibitem{Yarom:2009mw}
  A.~Yarom,
  JHEP {\bf 1004}, 024 (2010).
  [arXiv:0912.2100 [hep-th]].


\bibitem{Gibbons:1976ue}
  G.~W.~Gibbons, S.~W.~Hawking,
  Phys.\ Rev.\  {\bf D15}, 2752-2756 (1977).


\bibitem{Son:2002sd}
  D.~T.~Son and A.~O.~Starinets,
  JHEP {\bf 0209}, 042 (2002).


%

\bibitem{Policastro:2002tn}
  G.~Policastro, D.~T.~Son and A.~O.~Starinets,
  JHEP {\bf 0212}, 054 (2002)
  [arXiv:hep-th/0210220].



\bibitem{Hong:2010at}
  J.~Hong, D.~Teaney,
  Phys.\ Rev.\  {\bf C82}, 044908 (2010).
  [arXiv:1003.0699 [nucl-th]].




\bibitem{Starinets:2002br}
  A.~O.~Starinets,
  Phys.\ Rev.\  {\bf D66}, 124013 (2002).
  [hep-th/0207133].

\bibitem{Nunez:2003eq}
  A.~Nunez, A.~O.~Starinets,
  Phys.\ Rev.\  {\bf D67}, 124013 (2003).
  [hep-th/0302026].

\bibitem{Myers:2007we}
  R.~C.~Myers, A.~O.~Starinets, R.~M.~Thomson,
  JHEP {\bf 0711}, 091 (2007).
  [arXiv:0706.0162 [hep-th]].

\bibitem{Berti:2009kk}
  E.~Berti, V.~Cardoso, A.~O.~Starinets,
  Class.\ Quant.\ Grav.\  {\bf 26}, 163001 (2009).
  [arXiv:0905.2975 [gr-qc]].




\bibitem{Springer:2010mf}
  T.~Springer, C.~Gale, S.~Jeon {\it et al.},
  Phys.\ Rev.\  {\bf D82}, 106005 (2010).
  [arXiv:1006.4667 [hep-th]].




\bibitem{Csaki:2006ji}
  C.~Csaki, M.~Reece,
  JHEP {\bf 0705}, 062 (2007).
  [hep-ph/0608266].



\bibitem{Gursoy:2007cb}
  U.~Gursoy, E.~Kiritsis,
  JHEP {\bf 0802}, 032 (2008).
  [arXiv:0707.1324 [hep-th]].

\bibitem{Gursoy:2007er}
  U.~Gursoy, E.~Kiritsis, F.~Nitti,
  JHEP {\bf 0802}, 019 (2008).
  [arXiv:0707.1349 [hep-th]].





\bibitem{Klebanov:1999tb}
  I.~R.~Klebanov and E.~Witten,
  Nucl.\ Phys.\  B {\bf 556}, 89 (1999)
  [arXiv:hep-th/9905104].







\end{thebibliography}
\end{document}